# Non-negative matrix factorization based subband decomposition for acoustic source localization

S. Shon, S. Mun, D. Han and H. Ko

A novel Non-negative Matrix Factorization (NMF) based subband decomposition in frequency-spatial domain for acoustic source localization using a microphone array. The proposed method decomposes source and noise subband and emphasizes source dominant frequency bins for more accurate source representation. By employing NMF, we extract Delay Basis Vectors (DBV) and their subband information in frequency-spatial domain for each frame. The proposed algorithm is evaluated in both simulated noise and real noise with a speech corpus database. Experimental results clearly indicate that the algorithm performs more accurately than other conventional algorithms under both reverberant and noisy acoustic environments.

*Introduction:* Acoustic source localization has been an active research area with applications in a variety of fields and it has become an important topic in acoustic based applications. Time Delay Estimation (TDE) between two or more microphone signals can be used as a mean for source localization. Generalized Cross-Correlation (GCC) is the most commonly used TDE approach.

In this paper, we propose a decomposition of signal and noise subbands based on Non-negative Matrix Factorization (NMF) and GCC. Using the decomposed signal subband information, the source dominant frequency bins can be emphasized by spectral weighting. A TDE algorithm based on the proposed subband decomposition approach outperforms conventional GCC algorithms and other TDE algorithm such as Adaptive Eigenvalue Decomposition (AED) [1] and other spectral weighting method such as Cross-Power Spectrum (CPS) [2] and local-Peak-Weighted (LPW) [3] in reverberant and noisy environments. The proposed approach exhibits conceptual similarity to the Multiple Signal Classification (MUSIC) algorithm [4]. It decomposes the cross-correlation matrix of the multichannel signals into signal and noise subspaces using eigenvalue decomposition. It was developed originally as a direction-of-arrival (DOA) estimation technique for narrowband signals, and there are many variants. Although there are subspace techniques, such as the MUSIC method, that are applicable to wideband signals, theoretically, they cannot be used for coherent source localization such as acoustic environment with reverberations.

*Proposed subband decomposition:* Consider that the $M^{th}$ channel microphone input signal is $x_m(t)$ and its Short Time Fourier Transform (STFT) is $X_m(t)$, then the GCC with PHAse Transform (PHAT) of the $l^{th}$ and the $q^{th}$ microphone signal is

$$R_{lq}(\theta) = \frac{1}{2\pi} \int_{-\infty}^{\infty} \Psi_{lq}(\omega) X_l(\omega) X_q^*(\omega) e^{j\omega\theta} d\omega \qquad (1)$$

where $\Psi_{lq}$ denotes a PHAT weight function as $\Psi_{lq}(\omega) \equiv 1/|X_l(\omega)X_q^*(\omega)|$. Note that $\theta$ is azimuth when the Time Delay Of Arrival of the $l^{th}$ and the $q^{th}$ microphones is $\tau$ as $\theta = \sin^{-1}(\gamma\tau/d)$ where $d$ is the distance between the $l^{th}$ and the $q^{th}$ microphones, and $\gamma$ is the speed of sound.

Since STFT is designed for a discrete signal, frequency $\omega$ should be a discrete value, i.e., $\omega_k = 2\pi(k/N)$, where $N$ is the length of the frame and $k$ denotes the frequency bin index. Therefore, for calculating GCC-PHAT corresponding to each frequency bin, (1) can be rewritten as

$$R_{lq}(\theta, \omega_k) = \frac{1}{K} \Psi_{lq}(\omega_k) X_l(\omega_k) X_q^*(\omega_k) e^{j\theta\omega_k} \qquad (2)$$

Using (2), we show some examples of source localization in a single frame. For clean signals, we can see clear large amplitudes to source directions in all frequency bins as in Fig. 1 (a). However, when there is noise, the source signal is corrupted as shown in Fig. 1 (b). For more accurate and robust source localization, we utilize the NMF theory to decompose the source and noise subbands and accentuate the source dominant frequency bins.

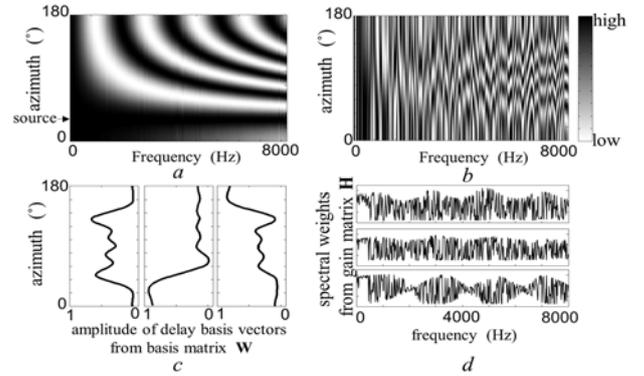

**Fig. 1** *Example of GCC-PHAT amplitude by each frequency and the proposed subband decomposition when the source is at 30˚.*

*a* GCC-PHAT *w*hen noise is absent: Clear large amplitude to source directions. Spatial aliasing cannot be avoided when *d* is larger than λ/2= *γ* /2*f* where λ is the wavelength of the signal frequency *f*
*b* GCC-PHAT when SNR = -5dB: Source signal is corrupted
*c* DBV **W**: the columns of **W** can be interpreted as DBV
*d* Subband information **H**: the rows of **H** are spectral weights corresponding to each DBV

NMF is a matrix factorization algorithm that decomposes non-negative matrix **V** in to a product of a non-negative basis matrix **W** and a non-negative gains matrix **H** as follows

$$\mathbf{V} \approx \mathbf{WH} \qquad (3)$$

where $\mathbf{V} \in \Re_+^{A \times B}$, $\mathbf{W} \in \Re_+^{A \times C}$ and $\mathbf{H} \in \Re_+^{C \times B}$ and $C<A,B$. For factorization, the Lee's approach was adopted in our method[5].

The most common usage of NMF is decomposing a spectrogram into spectral basis and its activation [6]. The non-negativity assumption of NMF algorithms leads to a parts-based representation. From this parts-based representation, it is successfully applied in many acoustic applications such as source separation and denoising problems.

Unlike common usage, we apply NMF to a GCC amplitude matrix **V** which represents the frequency-spatial domain as follows.

$$\mathbf{V} = \begin{bmatrix} R_{lq}(0,0) & \cdots & R_{lq}(0,K) \\ \vdots & \ddots & \vdots \\ R_{lq}(\pi,0) & \cdots & R_{lq}(\pi,K) \end{bmatrix} \qquad (4)$$

where the rows of **V** is determined by *K* which is overall number of frequency bins and the columns of **V** is determined by the size of azimuth resolution. As it is shown in (4), the spatial domain also can be regarded as TDOA domain. After NMF is applied in the GCC, we can get the set of Delay Basis Vectors (DBVs) **W** and their subband information **H**. According to the parts-based representation property of NMF, each DBV can be interpreted as a source DBV or noise DBV. Thus, the spectral weights contained by **H** decompose the subband information. Using this, we can achieve the main purpose of emphasizing source dominant frequency bins for more accurate and robust source localization.

After applying NMF to the **V** matrix, it is converged as

$$\mathbf{W} = \{\mathbf{w}_1, ..., \mathbf{w}_C\}^T, \ \mathbf{H} = \{\mathbf{h}_1, ..., \mathbf{h}_C\} \qquad (5)$$

$$\mathbf{h}_c = \{h_{c,0}, ..., h_{c,K-1}\}^T \qquad (6)$$

where *c* is basis index, *c*=1,2,…,*C*, and *C* is the total number of bases. To better understand the **W** and **H**, consider an example shown in Fig. 1 (c) and (d). We located the acoustic source at 30° under no reverberant but high noise (SNR=-5dB) environment. Finally, the proposed subband decomposition based spectral weighted GCC-PHAT for $\mathbf{h}_c$ can be expressed as

$$R_{lq}^{(c)}(\theta) = \frac{1}{K} \sum_{k=0}^{K-1} h_{c,k} \Psi_{lq}(\omega_k) X_l(\omega_k) X_q(\omega_k) e^{j\theta\omega_k} \qquad (7)$$

*Source DBV selection:* For applying the subband information in TDE as (7), we must determine which basis vector $\mathbf{w}_c$, *c*=1,2,…,*C*, is the source DBV. If we use a source DBV index *c*, $R_{lq}^{(c)}$ amplitude will show the highest peak at the source's existing azimuth, i.e. time delay, because the frequency bins that have the same time delay become emphasized. On the other hand, if we use a noise DBV index, $R_{lq}^{(c)}$ will not show clearly high peak because the noise signal has no coherent time delay.

For example, Fig. 2 shows an effect of the proposed algorithm with the same acoustic frame applied to that in Fig. 1. We first obtained the TDE result with a conventional algorithm under a clean environment as in Fig. 2 (a). Next, we obtained the TDE results of both conventional and proposed algorithms under SNR -5dB high noise environment as shown in Fig. 2 (b)~(e). It can be easily seen that the result using $\mathbf{h}_2$ is most similar to the clean environment result as in Fig. 2(a), while the others show significantly erroneous peaks. Therefore, determining the index of the source DBV can be done by finding the highest peak after applying all $C$ subband information as follows. The azimuth candidate $\theta_c$ corresponding to each DBV $c$ is

$$\theta_c = \underset{\tau_- < \tau < \tau_+}{\operatorname{argmax}} R_{lq}^{(c)}(\theta) . \qquad (8)$$

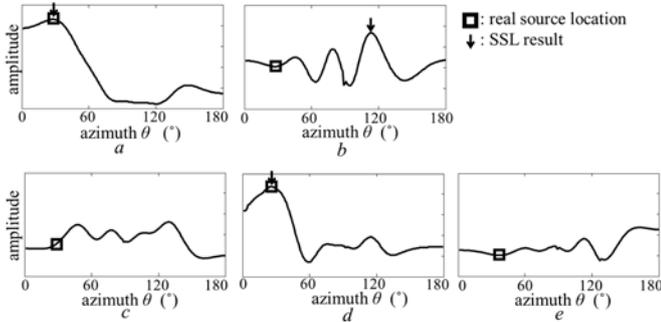

**Fig. 2** *TDE Examples*
*a* Conventional GCC-PHAT result when noise is absent
*b* Conventional GCC-PHAT result when SNR= -5dB
*c* Proposed result using (7) and $\mathbf{h}_1$ when SNR= -5dB
*d* Proposed result using (7) and $\mathbf{h}_2$ when SNR= -5dB
*e* Proposed result using (7) and $\mathbf{h}_3$ when SNR= -5dB

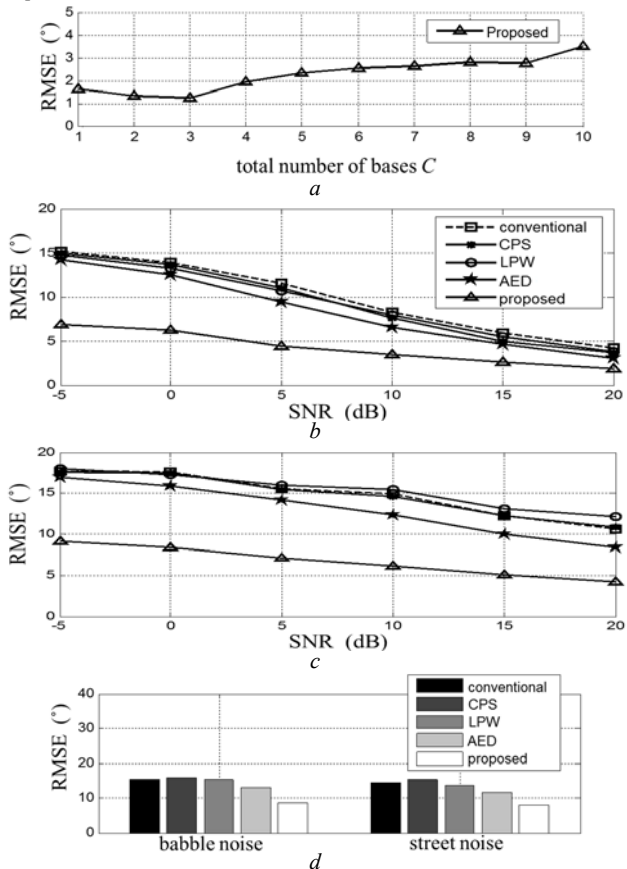

**Fig. 3** *TDE Performance evaluation using RMSE*
*a* With varying number $C$ of DBV on 20dB SNR with no reverb
*b* For varying SNR from -5dB to 20dB ($RT_{60}$=100ms)
*c* For varying SNR from -5dB to 20dB ($RT_{60}$=500ms)
*d* For real noise when SNR = -5dB

Then, we can find the azimuth of the source as

$$\theta_{source} = \max_c \left( R_{lq}^{(c)}(\theta_c) \right) \qquad (9)$$

The source DBV index $\beta$ is found by

$$\beta = \underset{c}{\operatorname{argmax}} \left( R_{lq}^{(c)}(\theta_c) \right). \qquad (10)$$

*Experiments:* The proposed algorithm was evaluated in a 3.5x4.5x2.5m³ simulated room environment. A pair of microphones with 10cm inter-spacing is located at the center of the room. The database was created with an acoustic source using noisy speech corpus (NOIZEUS) [7] which contains 720 sentences in the direction of 30°($\theta_s$) at 1.2m to the microphones. The database was composed of an additive Gaussian white noise at 16 kHz sampling rate. For reverberation, the room impulse response is generated by an image-source method. The length of STFT is 1024 using a Hamming window. We compared the proposed algorithm performance to the conventional GCC-PHAT, CPS and LPW, AED algorithms with Root Mean Square Error (RMSE).

To determine the optimal number of basis vectors, we first examine the performance of the proposed subband decomposition based spectral weighting of GCC-PHAT in terms of $C$, a number of DBV as in Fig. 3(a). It shows using too many DBVs leads to performance degradation. From the result, we set the number of DBV $C$ as 3.

In the next experiment, we fully evaluate robustness of the proposed subband decomposition based spectrally weighted GCC-PHAT. In both low and high reverberation environments as in Fig. 3(b) and (c), the conventional GCC-PHAT shows poor performance. The CPS, LPW and AED approaches show slight performance improvement. However, the proposed algorithm shows significant improvement in RMSE. Specifically, the proposed algorithm shows about 250% improvement in -5dB SNR environment compared to the conventional method. Interestingly, the LPW utilizing harmonics information shows poorer performance than the conventional algorithms. It can be interpreted that harmonics information was corrupted or not properly extracted in high reverberation environments.

We also conducted a performance evaluation under real noise as shown in Fig. 3(d). We used the 'babble place and 'street' noise with same database as [8]. SNR is -5dB. The proposed algorithm still seems to improve performance better than the other algorithms under real noise. From these analyses, the proposed subband decomposition based spectral weighting is shown to be robust to the noisy environment.

*Conclusion:* In this paper, we proposed the NMF-based subband decomposition for finding source dominant frequency bins and conducting TDE for achieving more robust acoustic source localization. The key feature of this paper is that we have applied NMF in the frequency-spatial domain for noise and source subband decomposition. From the experiments, it was shown that the proposed approach was more robust over other conventional algorithms in environments with low SNR and high reverberation.

S. Shon, S. Mun and H. Ko (*School of Electrical Engineering, Korea University, Seoul, Republic of Korea*)
✉ E-mail: hsko@korea.ac.kr

D. Han (*Assistant Secretary of Defence for Research and Engineering, United States Department of Defence, VA, USA*)